\documentclass[aps,prd,preprint,superscriptaddress]{revtex4}

\begin{document}

\title{The IGEX experiment revisited: a response to the critique of Klapdor-Kleingrothaus, Dietz, and Krivosheina}

\vskip1cm

\author{C. E. Aalseth}
\affiliation{Pacific Northwest National Laboratory, Richland, Washington 99352}
\author{F. T. Avignone III}
\affiliation{University of South Carolina, Columbia, South Carolina 29208}

\author{R. L. Brodzinski}
\affiliation{Pacific Northwest National Laboratory, Richland, Washington 99352}

\author{S. Cebrian}
\affiliation{University of Zaragoza, 50009 Zaragoza, Spain}

\author{E. Garcia}
\affiliation{University of Zaragoza, 50009 Zaragoza, Spain}

\author{D. Gonzales}
\affiliation{University of Zaragoza, 50009 Zaragoza, Spain}

\author{W. K. Hensley}
\affiliation{Pacific Northwest National Laboratory, Richland, Washington 99352}

\author{I. G. Irastorza}
\affiliation{University of Zaragoza, 50009 Zaragoza, Spain}

\author{I. V. Kirpichnikov}
\affiliation{Institute for Theoretical and Experimental Physics, 117259 Moscow, Russia}

 \author{A. A. Klimenko}
 \affiliation{Institute for Nuclear Research, Baksan Neutrino Observatory, 361309 Neutrino, Russia}

\author{H. S. Miley}
\affiliation{Pacific Northwest National Laboratory, Richland, Washington 99352}

 \author{A. Morales }
   \altaffiliation{Deceased}
 \affiliation{University of Zaragoza, 50009 Zaragoza, Spain}

\author{J. Morales}
\affiliation{University of Zaragoza, 50009 Zaragoza, Spain}

 \author{A. Ortiz de Solorzano}
 \affiliation{University of Zaragoza, 50009 Zaragoza, Spain}

\author{S. B. Osetrov}
\affiliation{Institute for Nuclear Research, Baksan Neutrino Observatory, 361309 Neutrino, Russia}

 \author{V. S.  Pogosov}
 \affiliation{Yerevan Physical Institute, 375036 Yerevan, Armenia}

\author{J. Puimedon}
\affiliation{University of Zaragoza, 50009 Zaragoza, Spain}
 
 \author{J. H. Reeves}
 \affiliation{Pacific Northwest National Laboratory, Richland, Washington 99352}
\affiliation{University of South Carolina, Columbia, South Carolina 29208}

\author{M. L. Sarsa}
\affiliation{University of Zaragoza, 50009 Zaragoza, Spain}

 \author{A. A. Smolnikov}
 \affiliation{Institute for Nuclear Research, Baksan Neutrino Observatory, 361309 Neutrino, Russia}

\author{A. S. Starostin}
\affiliation{Institute for Theoretical and Experimental Physics, 117259 Moscow, Russia}

 \author{A. G. Tamanyan}
 \affiliation{Yerevan Physical Institute, 375036 Yerevan, Armenia}

 \author{A. A. Vasenko}
 \affiliation{Institute for Theoretical and Experimental Physics, 117259 Moscow, Russia}

  \author{S. I. Vasiliev}
  \affiliation{Institute for Nuclear Research, Baksan Neutrino Observatory, 361309 Neutrino, Russia}

\author{J. A. Villar}
\affiliation{University of Zaragoza, 50009 Zaragoza, Spain}
 
\vskip1cm



\date{\today}

\vskip1cm

\begin{abstract}
This paper is a response to the article ``Critical View to'' the IGEX neutrinoless double-beta decay experiment...``published in Phys. Rev. D, Volume 65 (2002) 092007,'' by H. V. Klapdor-Kleingrothaus, A. Dietz, and I. V. Krivosheina, published as preprint hep-ph/0403056. The criticisms are confronted, and the questions raised are answered. We demonstrate that the lower limit quoted by IGEX, $T^{0\nu}_{1/2}(^{76}$Ge$)\geq 1.57\times 10^{25}$ y, is correct and that there was no ``arithmetical error''- as claimed in the ``Critical View'' article.
\end{abstract}


\maketitle


\section{INTRODUCTION}

The International Germanium Experiment (IGEX) operated with 3 detectors of $ \sim 700 $ gm fiducial mass and 3 detectors of $ \sim $ 2 kg fiducial mass each. They were fabricated from germanium isotopically enriched to $ 86\% $ in $ ^{76}$Ge. The total fiducial mass was 8.4 kg. They were mounted in ultra-low radioactive background cryostats electroformed from purified  CuSO$_{4}$ solution. The technical details were published in a number of earlier publications \cite{R1,R2,R3,R4}. In the paper in question by Aalseth et al., \cite{R5}, the IGEX collaboration reported their final results from  117 mole$\cdot$years of $ ^{76}$Ge data (mole$\cdot$years refers to moles of $ ^{76}$Ge). The numerical data from 2020 keV to 2060 keV were given and were used to place the lower bound: $ T_{1/2}^{0\nu}(^{76}$Ge$)\geq 1.57 \times 10^{25}$ y $(90\%\;$CL$)$. Earlier, the Heidelberg-Moscow (HM) collaboration reported the bound $ T_{1/2}^{0\nu}(^{76}$Ge$)\geq 1.9 \times 10^{25}$ y $(90\%)$ using a much larger body of data \cite{R6}.

At about the same time, a small subset of the HM collaboration published several papers claiming evidence for the observation of $ 0\nu \beta\beta $-decay of $ ^{76}$Ge \cite{R7,R8}. Their recent analysis of the data imply $ 6.8\times 10^{24}$ y $\leq T_{1/2}^{0\nu}(^{76}$Ge$)\leq 4.45 \times 10^{25}$ y $(4 \sigma\; $CL$)$, with the best fit value $ 1.19 \times 10^{25}$ y. A detailed critique of reference \cite{R7} was given in reference \cite{R9}. No mention of the claim of evidence \cite{R7,R8} was made in reference \cite{R5}, because reference \cite{R7} was published after the IGEX paper was completed, and at that time the IGEX collaboration was ignorant of the official position of the rest of the HM collaboration. The article entitled ``Critical View to ``The IGEX neutrinoless double-beta decay experiment''...published in Phys. Rev. D, Volume 65 (2002) 092007,'' is referred to throughout this paper as KKDK.  In KKDK the following statement appears in the abstract ``\emph{In view of the recently reported evidence for neutrinoless double beta decay (references given) it is particularly unfortunate that the IGEX paper is rather incomplete in its presentation}.'' The final IGEX paper was a brief statement of the final results with no intention of repeating the many details published earlier \cite{R1,R2,R3,R4}. Here we address the questions raised by KKDK, and are not concerned with the claim of evidence \cite{R7,R8}.

The statement is made in KKDK: ``\emph{The paper does not give sufficient detail on the history, quality, stability and run time of the detectors.  Also, for example the small ``duty cycle'' of the experiment is not explained.}''

The IGEX experiment was a research and development activity as well as an experiment.  It took several years to develop the technique allowing a field-effect transistor to be located within a few cm of the detector contact while still maintaining low background.  Keeping this distance to a minimum resulted in the excellent pulse shapes discussed in detail by Gonzales et al. \cite{R10}. In addition, the large crystals were grown one at a time to preserve the valuable isotopically enriched material.  

The first three detectors each had $ \sim $ 700 gm fiducial mass and were operated with one each in the Homestake gold mine, the Canfranc Tunnel Underground Laboratory in Spain, and the Baksan Neutrino Observatory in Russia, to evaluate the conditions at these sites. Considering overburden, location, and available space, the Canfranc Laboratory in Spain was ultimately chosen as the experimental site. Three larger detectors ($\sim$ 2 kg) were produced one at a time, tested, and eventually operated in Canfranc where operational conditions and space were excellent until the construction of a road tunnel parallel to the laboratory tunnel began. The laboratory tunnel was used to remove the excavated rock causing serious interruptions  that resulted in a very poor duty cycle and eventually the termination of the experiment. All the available data from Homestake, Baksan (where all three smaller detectors were eventually operated), and Canfranc were used in the analysis.  The details above are not of scientific interest and accordingly were not included in \cite{R5}. 

It is stated in KKDK, ``\emph{The background reached in the experiment is not even mentioned}.'' The background for the entire data set is trivially computed from Table II of reference \cite{R5}.  There were 69.9 counts in the 40-keV region from 2020 keV to 2060 keV in 10.14 kg$\cdot$years (all Ge). This results in 0.17 counts keV$^{-1}$ kg$^{-1}$ y$^{-1}$ in the data set, which represents $ 117\;(^{76}$Ge mole$\cdot$years) of data, $ 45\% $ of which was subjected to pulse shape discrimination (PSD). The background in the same 40-keV region, computed only from the $45\%$ of the PSD analyzed data, is 0.10 counts keV$^{-1}$ kg$^{-1}$ y$^{-1}$ \cite{R3}. In the introduction, KKDK also states, ``\emph{No analysis of  background lines has been published, and no Monte-Carlo simulations of the background is presented}.'' In the IGEX paper, a full spectrum could have been presented; however, all of the needed information could be obtained from the compressed 10 keV per channel spectrum given in reference \cite{R4}. The need for presenting Ge detector data in 0.36-keV bins, when the energy resolution is 4 keV is a matter of opinion. Many IGEX Monte-Carlo simulations have been done; however, none were needed in the analysis of the final IGEX data.  To obtain the lower limit of $ T^{0\nu}_{1/2}$, no corrections for the background were made aside from the PSD elimination of ``obviously'' multi-interaction-site events. It was assumed that nothing was known about the events in the interval 2020-2060 keV except what could be determined from PSD.  In a case making a claim that a peak at $ |Q(\beta\beta)| \simeq $ 2039 keV is due to $ 0\nu \beta\beta $ decay, it would be necessary to make an exhaustive investigation of all possible ramifications of background, as in the case of references \cite{R6,R7}. 

\section{PULSE SHAPE ANALYSIS}

In KKDK it is stated: \emph{``The method of pulse shape analysis (PSA) used in that paper seems not yet to be technically mature. It makes among others use of a visual determination of the shape of the pulses (references). This casts doubt on the reliability of the background determination.''} This doubt is a personal opinion with nothing to support it. In fact, a very complete study of the IGEX pulse shape analysis has been published by Gonzales et al., (the IGEX collaboration) \cite{R10}. The short distance (a few cm) between the gate of the field-effect-transistor and the detector contact resulted in single-site and multi-site pulses of the quality of those shown in references  \cite{R2,R3,R4,R10}. The pulse shapes of the IGEX PSD, in fact, compare very favorably to those of the HM experiment shown in reference \cite{R11}. This achievement of the IGEX collaboration was a non-trivial task and resulted in delays and the loss of detector operating time mentioned in KKDK. A total of only 32 pulses in the interval 2020-2060 keV were clearly identified visually to be multi-site. Three independent teams performed a triple blind selection, and only pulses agreed to by all three teams were removed from the data. Later, the techniques published by Gonzales et al. confirmed the validity of this procedure \cite{R10}. 

\section{STASTICAL ANALYSIS AND BACKGROUND}

The background issue has already been addressed above. The background that cannot be obviously attributed to gamma rays , i.e., that left after PSD, is 0.17 counts  keV$^{-1}$ kg$^{-1}$ y$^{-1}$. The total rate in the 40-keV interval prior to PSD is 102 counts \cite{R10} or 0.24 counts keV$^{-1}$ kg$^{-1}$ y$^{-1}$. The background in the $ \sim $ 53 mole$\cdot$years of PSD corrected data is 0.10 counts keV$^{-1}$ kg$^{-1}$ y$^{-1}$. Recently, significant technical improvements over those discussed in reference \cite{R10} have been made by the members of the IGEX collaboration \cite{R12,R13}.

A very detailed discussion of the apparatus, experimental procedures, and sources of background was given in a 1999 paper by Aalseth et al. \cite{R4}. The background spectrum in 10-keV bins shown in reference \cite{R4} gives the reader a clear view of the major background, while the sample pulse shapes shown are very descriptive of the quality of the IGEX PSD.

The lack of documentation of the statistical estimators used to extract the limit on $ T^{0\nu}_{1/2}$ was an oversight. Rather than to argue which method was most appropriate, the spectrum was presented with the statement, ``Readers can interpret the data in Table II as they wish.'' This, however, does not include arbitrarily choosing portions of the IGEX data to arrive at whatever results one wishes to, as was done by KKDK. It does not include analyzing only the 52.5 mole$\cdot$years of data, that were treated with PSD, without including the rest of the data. The analyses presented by KKDK are misleading and are completely incorrect. And, there are no ``arithmetic errors'' in the IGEX paper that lead to the published bound.

The data presented in Table II of the IGEX paper \cite{R5} represent 117 mole$\cdot$years of $ ^{76} $Ge data, corresponding to 8.89 kg$\cdot$years of $ ^{76} $Ge data, or 10.14 kg$\cdot$years using the total mass including $ 14\%\; ^{74}$Ge. This corresponds to: $ \ln 2\; N t = 4.88 \times 10^{25} $ y. The $ 90\% $ confidence limit of c $ < 3.1 $ possible $ 0\nu \beta\beta $-decay candidates was obtained first using a statistical estimator published by R. M. Bartlett et al. \cite{R14} and recommended by the particle data group. The same result was obtained using a standard maximum likelihood analysis. Finally, KKDK suggests that the IGEX authors should have used the ``Unified approach to the classical statistical analysis of small signals,'' by Feldman and Cousins \cite{R15}. It was pointed out earlier that the use of this technique can yield results dependent on the selection of the width of the region of interest chosen \cite{R16}. Nevertheless, if we choose the interval from 2034 keV to 2044 keV, which includes the entire expected $ \beta\beta $-decay peak, the expected background is 17.2. The number of events in this energy interval in the IGEX data is 9.6. Table V of Feldman and Cousins gives c = 3.00 $(90\%\; $CL$)$ for 10 events and 15 expected background events. Therefore, using the above numbers, the Feldman and Cousins technique leads to a bound less than c = 3.00 and a half-life greater than  $ 1.57 \times 10^{25}$ y, which is less conservative than our limit. The half life is computed on the basis of mole$\cdot$years of $ ^{76}$Ge data, whereas we compute the background on the basis of kg$\cdot$y of total Ge data as is the usual practice. 

\section{THE EFFECTIVE $ \nu $ MASS}

This section of KKDK begins with: ``\emph{Starting from their incorrectly determined half-life limit the authors claim a range of effective neutrino mass of  (0.33 - 1.35) eV}.'' In one case, KKDK selected only the 52.51 mole$\cdot$years of our data that had been subjected to PSD and obtained $ T_{1/2}^{0\nu} > 7.1 \times 10^{24}$ y using the maximum number of counts, 3.1, from the entire 117 mole$\cdot$years of data. This is erroneous and unjustified. In another case, KKDK also decided to arbitrarily use the entire IGEX data  set prior to PSD selection. From this they obtained a bound of  $ T_{1/2}^{0\nu} > 1.1 \times 10^{25}$ y. There is no scientific justification for selecting only PSD corrected data on one hand and totally ignoring the PSD corrected data on the other hand. The only correct treatment of the IGEX data is to  include the entire data set given in Table II of reference \cite{R5}. One could argue that there may be better ways of analyzing the complete data set; however, arbitrarily selecting part of it is not one of them.

The heading of the fourth column of their Table I stated: ``\emph{$ <m_{\nu}> eV $ from: our (conservative) analysis of Aalseth et al. data $ (90\%\; c.l.)\; 0.5\times 10^{25}\; $years,}'' is very misleading, because that half-life is erroneously derived by KKDK from our data as discussed above. 

\section{NUCLEAR STRUCTURE}

This section of KKDK begins:``\emph{The discussion of nuclear structure and matrix elements is incomplete and seems superficial. It ignores recent work (after 1996)}.'' This statement is correct. Table I of Aalseth et al. \cite{R5} was only meant to demonstrate the large discrepancy between models of the same technique. That shortcoming is corrected in this section. In Table I we give a more complete list of theoretical calculations published since 1996 \cite{R17,R18,R19,R20,R21,R22,R23,R24,R25,R26,R27} without commenting on them.  

To bring this discussion up to date, very recent calculations by two well known groups are discussed. In  a recent paper by Rodin, Faessler, Simkovic, and Vogel, a new approach was introduced \cite{R28}. In this paper it is stated: ``\emph{When the strength of the particle-particle interaction is adjusted so that the $ 2 \nu \beta\beta $-decay rate is correctly reproduced, the resulting $ M^{0\nu}$ values become essentially independent on the size of the basis, and on the form of different realistic nucleon-nucleon potentials. Thus one of the main reasons for variability of the calculated $ M^{0\nu} $ within these methods is eliminated}.'' These results are used to compute the limits on  $ \langle m_{\nu}\rangle $ from the IGEX data. 
\begin{equation}
\langle m_{\nu}\rangle = [|M^{0\nu}|(G^{0\nu} T_{1/2}^{0\nu})^{1/2}]^{-1}.
\end{equation}
In the above, M$^{0\nu}(^{76}$Ge$) = 2.40 \pm 0.07 $ for Renormalized Quasi-Particle Random-Phase Approximation (RQRPA), and M$^{0\nu}(^{76}$Ge$) = 2.68 \pm 0.06 $ for QRPA, with G$^{0\nu} = 0.30 \times 10^{-25}$ y$^{-1}$ eV$^{-2}$. This results in the following bounds: $ \langle m_{\nu}\rangle < 0.63$ eV (RQRPA), and $ \langle m_{\nu}\rangle < 0.56 $ eV (QRPA) for T$^{0\nu}_{1/2}(^{76}$Ge$)> 1.57\times 10^{25}$ y. 

There is no reason to constrain one's analysis to the matrix element calculations listed in KKDK. In fact the authors of references \cite{R18,R19,R20} and \cite{R23} quoted in KKDK were involved in the recent calculations of Rodin et al. \cite{R28}. Why should their earlier work be considered as current? Results similar to those of Rodin et al. were obtained by Civiterese and Suhonen \cite{R29}.  Table I of KKDK is also out of date. Recently, Engel and Vogel \cite{R30} give a critique of reference \cite{R28}, however, the basic conclusions of reference \cite{R28} remain intact.

\section{CONCLUSION}

In the conclusion of KKDK it states: ``\emph{the IGEX paper - apart from the too high half-life limits presented, as a consequence of an arithmetic error - is rather incomplete in its presentation}.'' It is clearly shown above that there was absolutely no arithmetic error. The analysis of the published IGEX  data presented in KKDK is not legitimate. To obtain a much shorter bound on the half-life, they arbitrarily analyzed two $ \sim $ halves of the data separately. Instead of having $ 4.88 \times 10^{25} $ y in the numerator $ (\ln 2 N t)$ they used $ 2.2 \times 10^{25} $ y. Yet they used the $(90\%)$ CL upper limit on the number of counts under the peak, obtained by IGEX from all of the data. In another analysis, they ignore the fact that 52.51 mole$\cdot$years were corrected with PSD and treat the complete uncorrected data set. Naturally, the lower limits on $ T^{o\nu}_{1/2} (^{76}$Ge$)$ obtained by these completely unjustified procedures are shorter than that obtained from properly analyzing the complete data set given in Table II of reference \cite{R5}.

The IGEX collaboration maintains its position that a proper analysis of the complete body of IGEX data results in a lower bound, $ T_{1/2}^{0\nu}\geq 1.57 \times 10^{25}$ y, and that it would not vary significantly from this value by the application of other appropriate statistical estimators. Several other estimators give similar results.

Finally, if the article by Aalseth et al. \cite{R5} relied too heavily on earlier IGEX publications, it is hoped that now unanswered questions have been adequately addressed.

\

\begin{table}[h]
 
  \caption{$ F_{N}\equiv G^{0\nu}| M^{0\nu}_{f} - (g_{A}/g_{V})^{2}M^{0\nu}_{GT}|^{2}$ of calculations after 1996. The effective Majorana mass of the electron neutrino, $ \langle m_{\nu}\rangle$, is given for $T^{0\nu}_{1/2} (^{76}Ge) = 1.57 \times 10^{25}\; y $.}

 \centering
\begin{tabular}{|c|c|c|}
  \hline
  $ F_{N}(y^{-1})$ & $ \langle m_{\nu}\rangle e V $ & Reference \\
  \hline
  $ 1.90 \times 10^{-14}$ & 0.94 & [17] \\
  $ 1.42 \times 10^{-14}$ & 1.09 & [18] \\
  $ 7.33 \times 10^{-14}$ & 0.48 & [18] \\
  $ 2.75 \times 10^{-14}$ & 0.78 & [19] \\
  $ 1.33 \times 10^{-13}$ & 0.35 & [20] \\
  $ 8.29 \times 10^{-14}$ & 0.45 & [21] \\
  $ 8.27 \times 10^{-14}$ & 0.45 & [22] \\
  $ 6.19 \times 10^{-14}$ & 0.51 & [23] \\
  $ 2.11 \times 10^{-13}$ & 0.29 & [23] \\
  $ 1.16 \times 10^{-13}$ & 0.38 & [24] \\
  $ 5.22 \times 10^{-14}$ & 0.56 & [25] \\
  $ 1.21 \times 10^{-14}$ & 1.17 & [26] \\
  $ 1.85 \times 10^{-14}$ & 0.94 & [26] \\
  $ 3.63 \times 10^{-14}$ & 0.67 & [26] \\
  $ 6.50 \times 10^{-14}$ & 0.51 & [26] \\
  $ 7.57 \times 10^{-14}$ & 0.46 & [27] \\
  \hline
\end{tabular}\label{t1}
\end{table}
\vskip0.5cm

\end{document}